\documentclass[twocolumn,showpacs,preprintnumbers,amsmath,superscriptaddress,amssymb,prb]{revtex4-1}
\usepackage{graphicx}
\usepackage{dcolumn}
\usepackage{bm}
\usepackage{booktabs}

\begin{document}

\title{Nonlinear Optical Properties of Transition Metal Dichalcogenide
 MX$_2$ (M = Mo, W; X = S, Se) Monolayers and Trilayers from First-principles Calculations}
\author{Chung-Yu Wang}
\address{Department of Physics and Center for Theoretical Sciences, National Taiwan University, Taipei 10617, Taiwan}
\author{Guang-Yu Guo}
\email{gyguo@phys.ntu.edu.tw}
\address{Department of Physics and Center for Theoretical Sciences, National Taiwan University, Taipei 10617, Taiwan}

\date{\today}

\begin{abstract}
Due to the absence of interlayer coupling and inversion symmetry, transition metal dichalcogenide (MX$_2$)
semiconductor monolayers exhibit novel properties that are distinctly different from their
bulk crystals such as direct optical band gaps, large band spin splittings, spin-valley coupling, piezoelectric and 
nonlinear optical responses, and thus have promising applications in, e.g., opto-electronic and spintronic
devices. Here we have performed a systematic first-principles study of the second-order nonlinear optical
properties of MX$_2$ (M = Mo, W; X = S, Se) monolayers and trilayers within the density functional 
theory with the generalized gradient approximation plus scissors correction.
We find that all the four MX$_2$ monolayers possess large second-order optical susceptibility $\chi^{(2)}$ 
in the optical frequency range and significant linear electro-optical coefficients in low frequency limit,
thus indicating their potential applications  in non-linear optical devices and electric optical switches.
The $\chi^{(2)}$ spectra of the MX$_2$ trilayers are overall similar to the corresponding MX$_2$ monolayers,
{\it albeit} with the magnitude reduced by roughly a factor of 3.
The prominent features in the $\chi^{(2)}$ spectra of the MX$_2$ multilayers
are analyzed in terms of the underlying band structures and optical dielectric function,
and also compared with available experiments.

\end{abstract}

\pacs{73.21.Ac, 78.20.Jq, 78.67.Pt, 42.70.Mp}

\maketitle

\section{Introduction}

Structurally, layered transition metal dichalcogenides of general formula MX$_2$ form a coherent family of compounds, 
where the metal (M) atoms are each coordinated by six chalcogen (X) atoms and each layer (monolayer) of the crystal is made up
of a two-dimensional (2D) hexagonal array of M atoms sandwiched between the similar arrays of X atoms.\cite{wil69}
The MX$_2$ sandwich monolayers are binded together by the weak van der Waals forces. They are highly
anisotropic and have been known as quasi-2D materials. Mechanically, therefore, they could be easily cleaved to prepare 
thin films\cite{wil69}, similar to graphite. Physically, however, the compounds have widely different electrical and optical 
properties and also host a number of interesting phenomena such as charge density wave and superconductivity phase transitions
(see, e.g., Refs. ~\onlinecite{wil69,bea76,bea79,guo86} and references therein). Chemically, the layered materials
may be intercalated by Lewis bases such as alkali metals and organic molecules (see, e.g., Refs. ~\onlinecite{lia86,guo87}
and references therein), often resulting in pronounced changes in their physical properties.
Therefore, the compounds had been under intensive investigations for nearly three decades since early 1960's.

In a recent optical experiment, a MoS$_2$ crystal was found to exhibit an indirect to direct band gap transition when
it is thinned down to a monolayer (ML).\cite{mak10} This discovery has triggered  
a growing renewed interest in the MX$_2$ semiconductors, {\it albeit} in their ML forms, because these MX$_2$ semiconductor
MLs exhibit fascinating properties that their bulk crystals do not have. Structurally, the MX$_2$ MLs have two distinct
differences from the MX$_2$ crystals, namely, lack of interlayer interaction and broken spatial inversion symmmetry.
The former causes the MX$_2$ MLs to become semiconductors with a direct band gap of $\sim$2.0 eV\cite{aku12,hon13}.
Therefore, the MX$_2$ MLs are promising materials for, e.g., electro-optical devices with efficient light emission\cite{mak10}
and field effect transitors with high on-off ratios\cite{rad11}.
The broken inversion symmetry, on the other hand, makes the MLs to exhibit novel properties of fundamental and technological interest
such as band spin-splitting\cite{zhu11}, spin-valley coupling\cite{xia12} and  piezoelectric property\cite{due12}. 

Being direct band gap semiconductors with noncentrosymmetry, the MX$_2$ MLs are also expected to show
significant second-order nonlinear optical susceptibility [$\chi^{(2)}$], and thus to provide novel applications in optoelectronics such as
coherent control of valley- and spin-polarized currents\cite{cao12}. Indeed, second-harmonic (SH) generation in the MoS$_2$ MLs has been observed
in recent experiments\cite{nku13,mal13,yil13}, although the reported $\chi^{(2)}$ modulus under
810 nm laser illumination varies as much as three orders of magnitude. 
To interpret the measured $\chi^{(2)}(\omega)$ spectra and also to help search and design new MX$_2$ MLs
with better nonlinear optical properties, {\it ab initio} material specific calculations of the $\chi^{(2)}$ 
would be needed. However, in contrast to the recent extensive theoretical investigations of the electronic, transport and linear optical properties 
of the MX$_2$ MLs, only theoretical 
calculations of $\chi^{(2)}$ for the MoS$_2$ ML\cite{Gru14,Tro14} have been reported. 

In this work, we systematically investigate the second-order nonlinear optical susceptibility and
also the linear electro-optical coefficient of all the four MX$_2$ (M = Mo, W; X = S, Se) 
MLs and trilayers (TLs). Our main goal is to find out the features and magnitude of 
the SH generation and linear electro-optical coefficients of
the MX$_2$ MLs in order to foresee their potential 
applications in nonlinear optical and electro-optical devices such as SH generation,
sum-frequency generation, electro-optical switch, and electro-optical modulator.
We also investigate the effects of the interlayer interaction on the second-order nonlinear 
optical properties by performing the {\it ab initio} calculations for the MX$_2$ TLs.

The rest of this paper is organized as follows. In Sec. II, the theoretical approach and 
computational details are briefly described. In Sec. III, the calculated band structure, 
density of states and second-order nonlinear optical susceptibility of
the MX$_2$ MLs and TLs are presented. Finally, a summary is given in Sec. IV.

\section{Theory and computational method}
The crystal structure of the MX$_2$ MLs is illustrated in Fig. 1(a).
The transition metal M atom sits on ($a/3, 2a/3, 0$) in the central plane, sandwiched by the chalcogen X atoms
on ($2a/3, a/3, \pm z$). Here $a$ is the in-plane lattice constant and $z$ is the distance between
the X and M atomic planes (Fig. 1).
To examine the effect of the interlayer interaction on the electronic and optical properties of the MX$_2$
multilayers, we also consider the MX$_2$ TLs [see Fig. 1(b)]. Note that the MX$_2$ bilayers and indeed all
the even number MX$_2$ multilayers do not
exhibit the second-order nonlinear optical response because they possess the spatial inversion symmetry.
In the MX$_2$ TLs, the three M atoms are located at ($a/3, 2a/3, 0$) and ($2a/3, a/3, \pm h$) while
the six X atoms sit on ($a/3, 2a/3, -h \pm z$), ($2a/3, a/3, \pm z$), and ($a/3, 2a/3, h \pm z$).
In the present calculations, the slab-supercell approach is adopted and a large vacuum slab of more than 18 \AA\
that separate the neighboring slabs is added in the direction perpendicular to the atomic planes.
The experimental structural parameters of the corresponding bulk crystals~\cite{thb01,wjs86} (see Table I)
are used in the present calculations. The effective thickness $h$ of one MX$_2$ ML is simply taken
as half of the lattice constant $c$ of the MX$_2$ crystal.
The effective thickness of the MX$_2$ trilayers is 3$h$.
\begin{table}
\caption{Experimental structural parameters for the MX$_2$ monolayers:  In-plane lattice constant $a$,
$z$-coordinate ($z$) of the X atoms, and effective thickness $h$ of one MX$_2$ ML. $h$ is simply
taken as half of the lattice constant $c$ of the corresponding bulk MX$_2$.
}
\begin{ruledtabular}
\begin{tabular}{c c c c c}
 MX$_2$ & MoS$_2$\footnote[1]{Reference ~\onlinecite{thb01}.}
      & MoSe$_2$\footnotemark[1]
      & WS$_2$\footnote[2]{Reference ~\onlinecite{wjs86}.}
      & WSe$_2$\footnotemark[2] \\  \hline
 $a$ (\AA)  & 3.160 & 3.299 & 3.152 & 3.282 \\
 $2z$ (\AA)  & 3.172 & 3.338 & 3.142 & 3.341 \\
 $h$ (\AA)  & 6.147 & 6.469 & 6.162 & 6.481 \\
\end{tabular}
\end{ruledtabular}
\end{table}

\begin{figure}
\includegraphics[width=8cm]{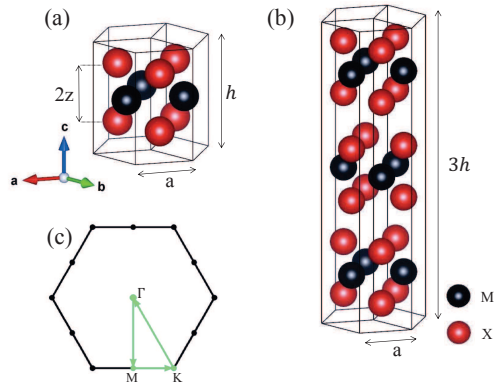}
\caption{\label{fig1}
Atomic structure of the MX$_2$ monolayers (a) and trilayers (b) as well as the associated Brillouin zone (c).}
\end{figure}

\subsection{Band structure calculation}

The present first-principles calculations are based on the density functional theory with the
generalized gradient approximation (GGA) of Perdew, Burke and Ernzerhof~\cite{jpp96}.
The accurate full-potential projector-augmented wave (PAW)
method~\cite{blo94}, as implemented in the VASP package~\cite{kre93}, is used.
A large plane-wave cut-off energy of 400 eV is used throughout.
The self-consistent band structure calculations are performed with a dense $k$-point grid of
20$\times$20$\times$1. For comparison, we also perform the same first-principles calculations 
for bulk MX$_2$ crystals. A $k$-point grid of 20$\times$20$\times$5 is used for the bulk calculations.

\subsection{Calculation of the optical properties}
In this work, the linear optical dielectric function and nonlinear optical susceptibility 
are calculated based on the linear response formalism with the
independent-particle approximation, as described before~\cite{Guo04,Guo05a,Guo05b}. 
The imaginary part [$\varepsilon''(\omega)$] of the dielectric function due
to direct interband transitions is calculated from the final self-consistent 
electronic band structure by using the Fermi golden rule~\cite{Guo04,Guo05a}.
The real part $[\varepsilon'(\omega)$] of the dielectric function is obtained 
from $\varepsilon''(\omega)$ by a Kramer-Kronig transformation
\begin{equation}
 \varepsilon'(\omega) = 1+\frac{2}{\pi}{\bf P} \int_0^{\infty}d\omega'
 \frac{\omega'\varepsilon''(\omega')}{\omega'^2-\omega^2}.
\end{equation}
Given the complex dielectric function ($\varepsilon'+i\varepsilon''$),
all other linear optical properties such as refractive index,
reflectivity and absorption spectrum can be calculated.

Following previous nonlinear optical calculations~\cite{Guo04,Guo05b}, the imaginary
part [$\chi''^{(2)}_{abc}(-2\omega,\omega,\omega)$] of the second-order optical susceptibility 
due to direct interband transitions is obtained from the self-consistent
electronic band structure by uising the expressions already given elsewhere~\cite{Guo04,Guo05b}.
The real part of the second-order optical susceptibility is then obtained from
$\chi''^{(2)}_{abc}$ by a Kramer-Kronig transformation
\begin{equation}
\chi'^{(2)}(-2\omega,\omega,\omega) = \frac{2}{\pi}{\bf P} \int_0^{\infty}d\omega'
 \frac{\omega'\chi''^{(2)}(2\omega',\omega',\omega')}{\omega'^2-\omega^2}.
\end{equation}
 
The linear electro-optic coefficient $r_{abc}(\omega)$ is connected to the second-order optical 
susceptibility $\chi_{abc}^{(2)}(-\omega,\omega,0)$ through the relation~\cite{jlp96}
\begin{equation}
\chi_{abc}^{(2)}(-\omega,\omega,0)=-\frac{1}{2}n_a^2(\omega)n_b^2(\omega)r_{abc}(\omega)
\end{equation}
where $n_a(\omega)$ is the refraction index in the $a$-direction. 
Therefore, in the zero frequency limit,
\begin{equation}
r_{abc}(0)=-\frac{2}{n_a^2(0)n_b^2(0)}\lim_{\omega\rightarrow 0}\chi_{abc}^{(2)}(-2\omega,\omega,\omega)
\end{equation}
Furthermore, for the photon energy $\hbar\omega$ well below the band gap the linear electro-optic coefficient
$r_{abc}(\omega)\approx r_{abc}(0)$ because $\chi_{abc}^{(2)}(-2\omega,\omega,\omega)$ and $n(\omega)$ are
nearly constant in the very low frequency region.\cite{Guo04,Guo05b}

In the present calculations, the $\delta$-function in the Fermi golden rule formulas\cite{Guo04,Guo05a,Guo05b}
is approximated by a Gaussian function
\begin{equation}
 \delta(x) \approx \frac{1}{\sqrt{\pi}\Gamma}e^{-x^2/\Gamma^2},
\end{equation}
with $\Gamma = 0.1$ eV.
To obtain accurate optical properties, the $k$-point grid used is much denser than that used in the self-consistent
band structure calculations (Sec. II.A).
We use a $k$-point grid of 130$\times$130$\times$1 for the MX$_2$ MLs and of 100$\times$100$\times$1 for
the MX$_2$ TLs.  Furthermore, to ensure that $\varepsilon'$ and also $\chi'^{(2)}$ calculated via Kramer-Kronig
transformation are accurate, at least twenty five energy bands
per atom are included in the present optical calculations. 
The unit cell volume $\Omega$ of the MX$_2$ MLs and TLs 
in the slab-supercell approach is not well defined. 
Here we use an effective unit cell volume $\Omega$ that is given by the area of
the in-plane unit cell times the effective thickness of the MX$_2$ ML ($h$) or TL ($3h$) (Fig. 1).
\begin{figure}
\includegraphics[width=8cm]{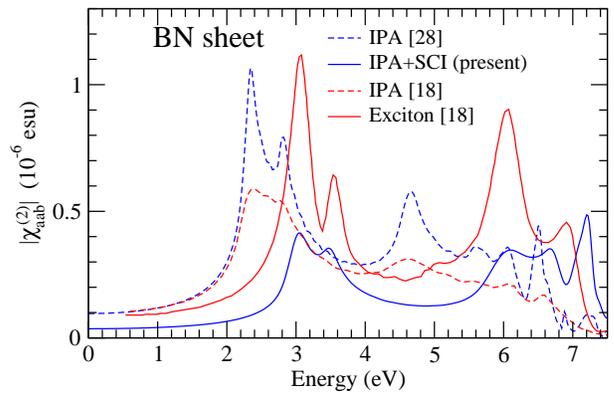}
\caption{\label{fig2}
Theoretical $|\chi^{(2)}|$ of the single BN sheet from the previous IPA calculations\cite{Guo05b,Gru14},
real-time propagation calculation which includes the exciton effect\cite{Gru14} and the present IPA+SCI calculation.
The discernable differences in $|\chi^{(2)}|$ between the two IPA calculations above 2.2 eV are
due to the much different $k$-point grids, namely, $40\times40\times1$ and $100\times100\times2$,
used in Refs. \onlinecite{Gru14} and \onlinecite{Guo05b}, respectively.
}
\end{figure}

In this work, the linear and nonlinear optical properties are calculated based on the
independent-particle approximation (IPA), i.e., the quasi-particle self-energy corrections and
excitonic effects were neglected. 
However, these many-body effects on the linear optical properties of 2D
systems such as SiC, MoS$_2$ and WSe$_2$ MLs\cite{Hsu11,Qiu13,Wan14} are especially pronounced due to 
quantum comfinement. Nevertheless, accurate {\it ab initio} calculations 
of the optical properties including the excitonic effect are usually extremely demanding computationally.
Indeed, it was recently demonstrated that a convincing agreement between experimental and theoretical 
absorption spectra could be achieved in the Bethe-Salpeter exciton approach only when 
several thousand $k$-points and tens of bands were included in the calculations.\cite{Qiu13}
Furthermore, {\it ab initio} calculations of nonlinear optical properties face another challenge
which comes from the complexity of the expression for the correlated nonlinear susceptibility 
in terms of the electronic band structure, and this makes the full {\it ab initio} calculations
within many-body perturbation theory impractical. Consequently, much simpler approaches such as
the real-time propagation\cite{Gru14} and semiempiral tight-binding-based model potential\cite{Tro14}
were adopted in recent calculations of the second-order nonlinear optical susceptibility 
for the MoS$_2$ ML. In this work, instead, we introduce the so-called scissors 
correction (SCI) to reduce the errors caused by the neglected many-body effects. 
This simple approach allows us to carry out a systematic investigation of the nonlinear optical
properties of all the four MX$_2$ materials in both the ML and TL structures,
using large numbers of $k$-points and conduction bands which are needed to
ensure that the theoretical results are numerically reliable. 
In Fig. 2, we display the $|\chi^{(2)}|$ of the single BN sheet calculated
within the IPA\cite{Guo05b,Gru14} and also the real-time approach to the excitonic effect\cite{Gru14}. 
It is clear from Fig. 2 that all the principal features in the $|\chi^{(2)}|$ spectrum
from the real-time approach\cite{Gru14} are more or less reproduced by the IPA calculation\cite{Guo05b}
except the red-shift of the peak energy positions. Furthermore, away from the excitonic resonances,
the values of $|\chi^{(2)}|$ from both approaches are rather close. For example, the  $|\chi^{(2)}(0)|$ 
values from Refs. \onlinecite{Guo05b} and  \onlinecite{Gru14} are 40.7 and 41.2 pm/V, respectively.
A scissors correction to the IPA with an energy shift of 1.34 eV brings the two spectra
in good agreement, {\it albeit}, with the $|\chi^{(2)}|$ magnitude from the present IPA+SCI being
less than half of that from the real-time approach (see Fig. 2). Note that the recent experimental
estimation\cite{yil13} of $|\chi^{(2)}|$ of $\sim$20.8 pm/V at 1.53 eV is nearly identical to
that of $\sim$20.7 pm/V from the present IPA+SCI calculation, while, in contrast, it is much smaller than
that from the real-time approach ($\sim$92 pm/V)\cite{Gru14} and the IPA calculation\cite{Guo05b} (68 pm/V). 

\section{Results and discussion}

\subsection{Band structures of MX$_2$ Monolayers}

The calculated band structures as well as total and site-decomposed densities of states of the four MX$_2$
MLs studied here are displayed in Fig. 3, and the corresponding band gaps are listed in Table II.
Figure 3 shows that all the four  MX$_2$ MLs are semiconductors with a direct band gap at the K symmetry point,
as found in previous optical experiments on the MoS$_2$ MLs~\cite{mak10}.
Furthermore, Table II suggests that the magnitude of the band gaps are in the visible light wavelengths and thus
the direct band gaps can be observed in photoluminescence experiments~\cite{ton13,hrg13}.
Therefore, all these four MX$_2$ MLs have promising potentials for electronic, optical and electro-optical devices.
We note that the calculated band structures of the MX$_2$ MLs (Fig. 3) are in good agreement with
previous GGA and local density approximation (LDA) calculations\cite{hon13,aku12}.

\begin{figure}
\includegraphics[width=8cm]{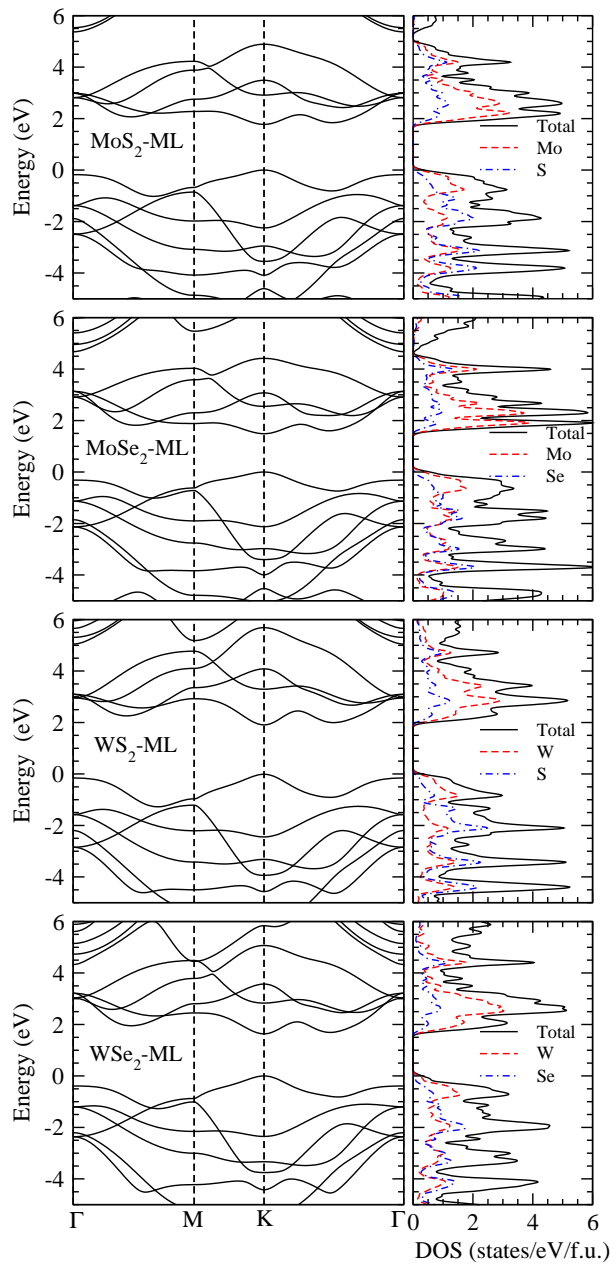}
\caption{\label{fig3}
Calculated band structures (left panels) and density of states (right panels) of the MX$_2$ MLs. 
All the four materials possess a direct band gap at the K-symmetry point.  
The top of the valence band is at 0 eV.}
\end{figure}

Since all the MX$_2$ MLs have the same crystalline structure and are isoelectronic, 
their electronic band structures are rather similar, as Fig. 3 shows. In particular, the left panels in Fig. 3
show that all of them have the M $d_{z^2}$-dominated top valence band with certain X $p_z$ component
and also the strongly X $p$-orbital and M $d$-orbital hybridized lower valence bands. Thus
the bonding of these compounds is mainly of covalent type. The lower conduction bands of the MX$_2$ MLs 
are M $d$-dominant bands with significant X $p$-orbital contributions.
Therefore, the optical transitions would be dominated by
the transitions from the valence states of chalcogen $p$-orbital and metal $d$-orbital hybrid 
to the conduction states of metal $d$ character.   

Nonetheless, there are minor differences among the band structures of the MX$_2$ MLs.
Table II shows that the band gap decreases as the S atoms are replaced by the Se atoms. On the other hand, 
when the chalcogen atoms are kept, the band gap becomes slightly larger if the Mo atoms are substituted by the W atoms.
Furthermore, for the MoS$_2$ and MoSe$_2$ MLs, there is a gap in the conduction band at around 5 eV while 
this gap is absent in the WS$_2$ and WSe$_2$ MLs (Fig. 3). Finally, the WS$_2$ ML has the largest band gap
while the MoSe$_2$ ML has the smallest one.

\begin{table}
\begin{ruledtabular}
\caption{Calculated and experimental band gap ($E_{g}$) and direct energy gap at K point ($E_{K}$)
of MX$_2$ MLs and TLs as well as bulk crystals. For the MLs, $E_{g} = E_{K}$.
$\Delta E_K^{ML}$ and $\Delta E_K^{TL}$ denote, respectively, the differences in the $E_{K}$ 
between the MLs (the.) and the bulks (exp.) as well as between the TLs (the.) and the bulks (exp.).
$\Delta E_g$ and $\Delta E_K^{bulk}$ represent, respectively, the differences in the band gap and
the energy gap ($E_{K}$) of bulk MX$_2$ crystals between the present calculations and previous 
experiments~\cite{bea79,bea76}. 
}
\begin{tabular}{c c c c c c}
       &             & MoS$_2$ & MoSe$_2$ & WS$_2$ & WSe$_2$ \\  
(a) MLs  &           &         &          &        &          \\ \hline
 $E_{g}$ (eV) & the. & 1.78 & 1.49 & 1.91 & 1.64 \\
 $\Delta E_K^{ML}$ (eV) & & 0.10 & 0.08 & 0.15 & 0.07 \\
        &           &         &          &        &          \\
(b) TLs  &           &         &          &        &          \\        
 $E_{g}$ (eV) & the. & 1.08 & 1.00 & 1.22 & 1.15 \\
 $E_{K}$ (eV) & the. & 1.72 & 1.43 & 1.85 & 1.55 \\
 $\Delta E_K^{TL}$ (eV) & & 0.16 & 0.14 & 0.21 & 0.16 \\
        &           &         &          &        &          \\
(c) bulk  &           &         &          &        &          \\        
 $E_{g}$ (eV) & the. & 0.87 & 0.82 & 0.99 & 0.95 \\
         & exp. & 1.29\footnote[1]{Experimental values from Ref. ~\onlinecite{bea79}.} & 1.10\footnotemark[1] &
                  1.30\footnote[2]{Experimental values from Ref. ~\onlinecite{bea76}.} & 1.20\footnotemark[2] \\
 $\Delta E_g$ (eV) & & 0.42 & 0.28 & 0.31 & 0.25 \\
 $E_{K}$ (eV) & the.& 1.69 & 1.40 & 1.83 & 1.53 \\
              & exp.& 1.88\footnotemark[1] & 1.57\footnotemark[1] & 2.06\footnotemark[2] & 1.71\footnotemark[2] \\
 $\Delta E_K^{bulk}$ (eV) & & 0.19 & 0.17 & 0.23 & 0.18

\end{tabular}
\end{ruledtabular}
\end{table}

\subsection{Second-order nonlinear optical susceptibility of MX$_2$ Monolayers}

\begin{figure*}
\includegraphics[width=16cm]{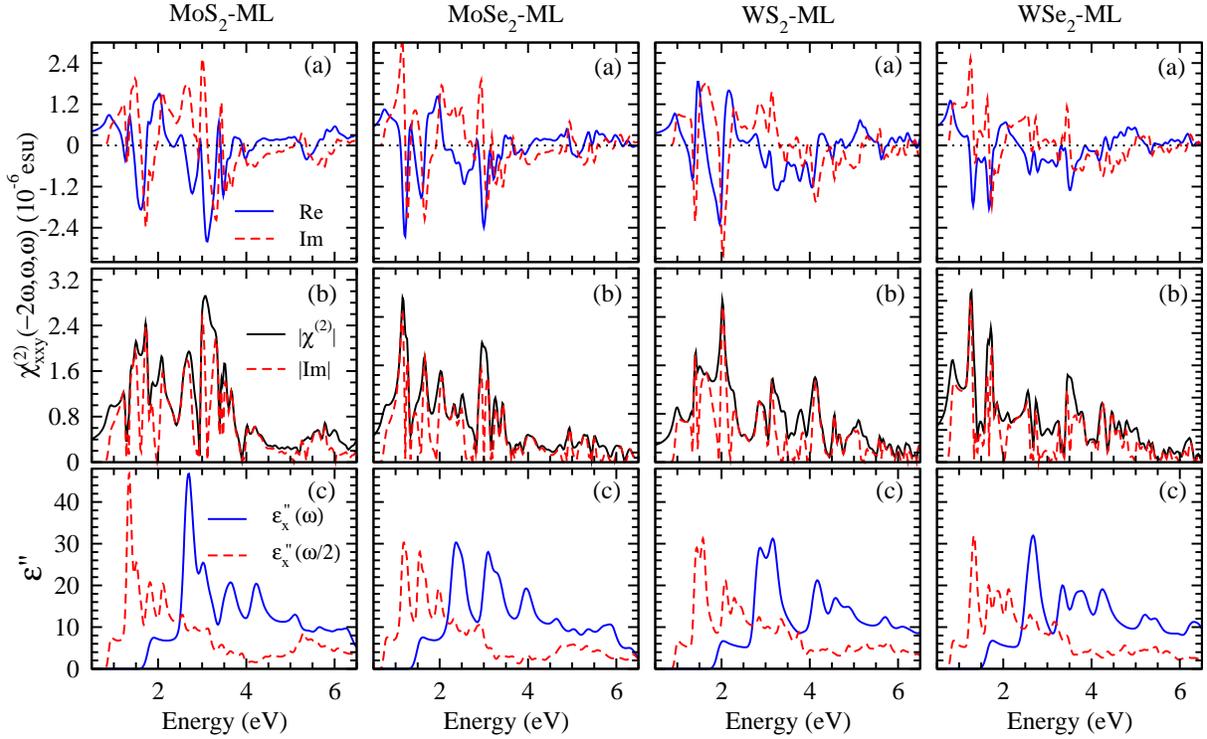}
\caption{\label{fig4}
(a) Real and imaginary parts as well as (b) the absolute value of the imaginary
part of the second-order susceptibility $\chi^{(2)}_{xxy}$ of the MX$_2$ MLs.
(c) The imaginary part of the dielectric function $\varepsilon''$ of the MX$_2$ MLs. }
\end{figure*}

\begin{table}
\caption{Calculated static refraction index ($n_x$), second-order optical susceptibility $\chi^{(2)}(0)$, 
$|\chi^{(2)}(1.53$ eV$)|$ and linear electro-optical coefficient $r_{xxy}$ of the MX$_2$ MLs (a)
and TLs (b) using the band structures without (IPA) and with (SCI) scissors correction.
The available experimental values (exp.) are also listed.
}
\begin{ruledtabular}
\begin{tabular}{ c c c c c c }
   &   & MoS$_2$ & MoSe$_2$ & WS$_2$ & WSe$_2$\\ 
(a) MLs & & & & & \\ \hline
 $n_x$ & IPA & 3.92 & 3.91 & 3.76 & 3.75 \\
       & SCI & 3.65 & 3.64 & 3.48 & 3.51 \\
 $\chi_{xxy}^{(2)}(0)$ (pm/V) & IPA & 141 & 170 & 125 & 177 \\
                              & SCI & 109 & 128 &  93 & 132 \\
$|\chi_{xxy}^{(2)}(1.53)|$ (pm/V) & IPA & 847 & 558 & 819 & 227 \\
                                  & SCI & 573 & 449 & 712 & 163\\
                            & exp. & 322\footnote[1]{Experimental value from Ref. ~\onlinecite{yil13}.},82\footnote[2]{Experimental value from Ref. ~\onlinecite{mal13}.} & & & \\
                            & exp.\footnote[3]{Experimental value from Ref. ~\onlinecite{nku13}.} & 10$^5$,5000\\
  $r_{xxy}(0) $(pm/V) & IPA & -1.19 & -1.45 & -1.25 & -1.79 \\
                   & SCI & -1.23 & -1.46 & -1.26 & -1.74 \\
       & & & & & \\
(b) TLs & & & & & \\           
 $n_x$ & IPA  & 3.96 & 3.93 & 3.79 & 3.78 \\
       & SCI  & 3.69 & 3.55 & 3.51 & 3.54 \\
 $\chi_{xxy}^{(2)}(0)$ (pm/V) & IPA & 49 & 58 & 43 & 58 \\
                              & SCI  &38 & 44 & 31 & 44 \\
$|\chi_{xxy}^{(2)}(1.53)|$ (pm/V) & IPA & 257 & 158 & 257 & 144 \\
                                  & SCI & 164 & 180 & 267 & 166 \\
                                  & exp.& 80\footnotemark[1],17\footnotemark[2] & & & \\
  $r_{xxy}(0) $(pm/V) & IPA & 0.39 & 0.43 & 0.40 & 0.35 \\
                      & SCI  & 0.40 & 0.43 & 0.40 & 0.35 \\
       & & & & & \\
(c) TLs vs. MLs & & & & & \\
  $\chi_{TL}^{(2)}(0)/\chi_{ML}^{(2)}(0)$ & IPA & 0.35 & 0.34 & 0.34 & 0.33 \\
                                          & SCI & 0.35 & 0.34 & 0.34 & 0.33 \\
 $\chi_{TL}^{(2)}/\chi_{ML}^{(2)}(1.53)$ & IPA & 0.30 & 0.28 & 0.31 & 0.63 \\
                                               & SCI & 0.28 & 0.40 & 0.39 & 1.02 \\
                                               & exp.& 0.25\footnotemark[1],0.21\footnotemark[2] & & 0.60\footnote[4]{Experimental value from Ref. ~\onlinecite{zen13}.} & 0.93\footnotemark[4] 
\end{tabular}
\end{ruledtabular}
\end{table}

Bulk MX$_2$ crystals have zero second-order nonlinear susceptibility since
their symmetry class is D$_{6h}$ with the spatial inversion symmetry. However, the MX$_2$ MLs
have the D$_{3h}$ symmetry without the inversion symmetry. Therefore, the MX$_2$ MLs 
would exhibit the second-order nonlinear optical response with nonzero susceptibility 
elements of $\chi^{(2)}_{xxy}$ = $\chi^{(2)}_{xyx}$ = $\chi^{(2)}_{yxx}$ = $-\chi^{(2)}_{yyy}$, 
as dictated by the D$_{3h}$ symmetry. 
Here subscripts $x$ and $y$ denote the two Cartesian coordinates in the MX$_2$ ML plane. 
Our theoretical results are consistent with this symmetry consideration, demonstrating that 
our numerical method and calculations are qualitatively correct.
The calculated real and imaginary parts as well as the modulus of the
imaginary part of $\chi^{(2)}_{xxy}(-2\omega,\omega,\omega)$ are shown in Fig. 4. 

It is well known that the band gaps from both the LDA and GGA calculations are usually smaller 
than that measured in optical experiments. For example, Table II shows that the calculated band gaps
of bulk MX$_2$ are smaller than the measured values by about 20$\sim$30 \%. 
It is clear from the Fermi golden rule formulas~\cite{Guo04} that the smaller the size of the energy gap
between the initial and final states on each $k$-point in the Brillouin zone, the larger the magnitude
of the second-order nonlinear susceptibility and dielectric function.
In other words, the optical calculations using a GGA band structure may overestimate 
the second-order nonlinear susceptibility and dielectric function.
To reduce this overestimation, we repeat the optical calculations using the scissors corrected
band structures. In the present scissors corrections, we use the energy differences ($\Delta E_K^{bulk}$) 
between the measured (by optical absorption) direct band gaps of bulk MX$_2$ crystals and calculated 
energy gaps of the MX$_2$ MLs at the $k$-point (see Table II) to shift the conduction bands upwards.  
The optical band gap of some MX$_2$ MLs has been measured by photoluminescence (PL) 
experiments~\cite{ton13,hrg13}. However, the PL measurement usually underestimates the band gap. 
On the other hand, a recent optical absorption experiment~\cite{mak10} showed that the band gap 
of the MoS$_2$ ML is the same as the direct band gap of bulk MoS$_2$ (1.88 eV)~\cite{bea76}. 
Therefore, we expect that the band gaps of the MX$_2$ MLs 
are close to the direct band gaps of the corresponding bulk MX$_2$ crystals. The real and imaginary 
parts as well as the absolute value of the imaginary part of $\chi^{(2)}_{xxy}(-2\omega,\omega,\omega)$ 
of the MX$_2$ MLs calculated from the scissors corrected band structures, are displayed in Fig. 5.
Figures 4 and 5 show that although the line shapes of the SH generation coefficient
and dielectric function from the two calculations are nearly identical, the magnitude of these optical
quantities from the scissors correction calculations gets reduced by about 25 \% and the peak positions
is shifted upwards by about $\Delta E_K^{bulk}$ (see also Table III).  
In the rest of this paper, we will concentrate mainly on the optical properties calculated with
scissors corrections which should give rise to more accurate optical quantities.

\begin{figure*}
\includegraphics[width=16cm]{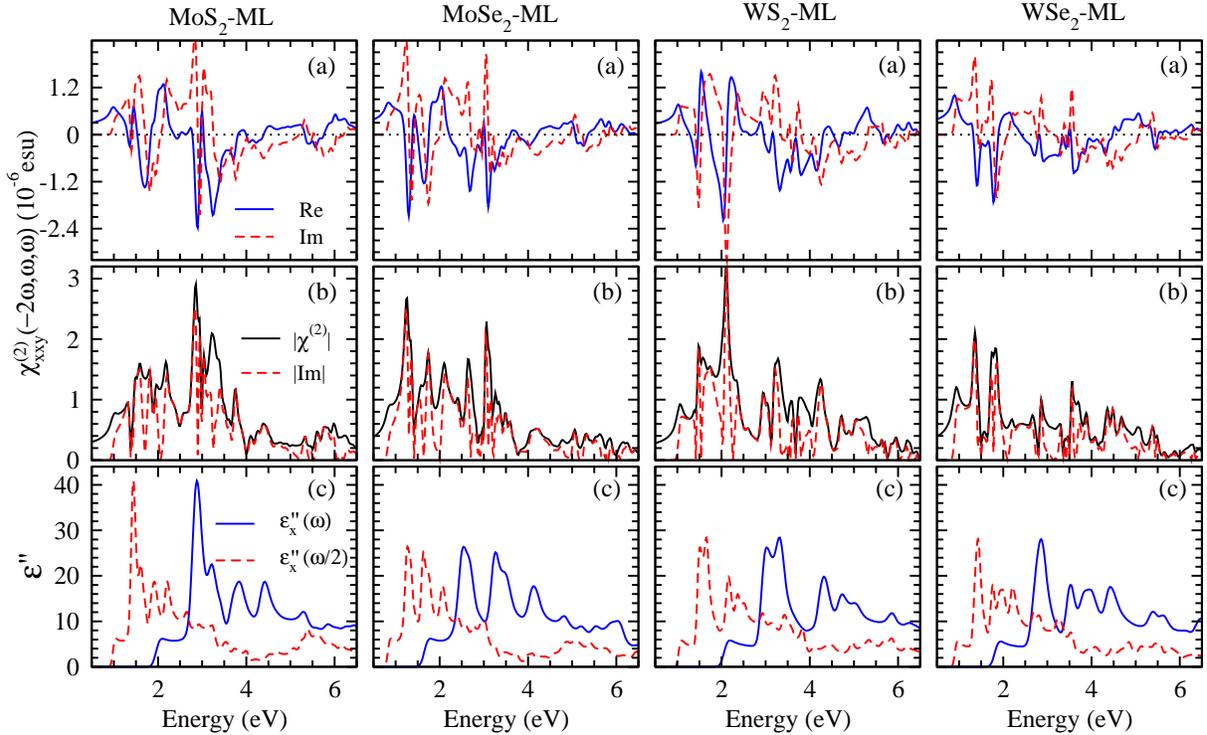}
\caption{\label{fig5}
(a) Real and imaginary parts as well as (b) the absolute value of the imaginary
part of the second-order susceptibility $\chi^{(2)}_{xxy}$ of the MX$_2$ MLs calculated
from the scissors corrected band structures. (c) Imaginary part $\varepsilon''$ of
the corresponding dielectric function of the MX$_2$ MLs. }
\end{figure*}

Figuer 5 indicates that the $\chi^{(2)}_{xxy}(-2\omega,\omega,\omega)$ 
of the MX$_2$ MLs are large in the entire range of optical photon energy, in the sense that they are 
comparable to that of GaAs~\cite{ber03}, an archetypical nonlinear optical semiconductor. 
We note that these SH susceptibilities are several times larger than that of the graphitic BN sheet\cite{Guo05b}. 
Furthermore, $\chi^{(2)}_{xxy}(-2\omega,\omega,\omega)$ of the MX$_2$ MLs is purely dispersive 
for photon energy being smaller than half of the direct band gap, because the absorptive part of $\chi^{(2)}_{xxy}$ 
becomes nonzero only for photon energy larger than half of the band gap [see Table II, Figs. 4(a) and 5(a)].
Table III also indicates that low frequency linear electro-optic coefficients of these monolayers are also large.
All these suggest that the MX$_2$ MLs may have application potentials in second order nonlinear optical devices 
and linear electro-optic modulators.

In general, the static SH susceptibility is small for a MX$_2$ ML with a large band gap.
Table III shows that this is indeed the case, except that 
the WSe$_2$ ML has the largest SH generation coefficient but does not have the smallest band gap. 
This may be explained by the fact that the fifth conduction band of the WSe$_2$ ML 
is lower than the MoSe$_2$ ML, and this may give rise to a larger static value via the Kramers-Kronig transformation.

To analyze the prominent features in the calculated $\chi^{(2)}(\omega)$ spectrum in a MX$_2$ ML, 
it is helpful to compare the magnitude of the imaginary part of $\chi^{(2)}(\omega)$ with the absorptive part 
of the corresponding dielectric function $\varepsilon''(\omega)$. Figures 4 and 5 show that the peaks in the $|$Im$[\chi^{(2)}(\omega)]|$
in the energy range from the absorption edge of $\varepsilon''({\omega/2})$ to the absorption edge of $\varepsilon''({\omega})$ 
can be correlated with the features in the $\varepsilon''({\omega/2})$ spectra, indicating that they are due to two-photon resonances. 
The peaks above the absorption edge of $\varepsilon''({\omega})$, on the other hand, can be related to the features in
either the $\varepsilon''({\omega/2})$ or $\varepsilon''({\omega})$ or both, suggesting that they can be caused by 
both double-photon and single-photon resonances. Due to the contributions from both one and two photon resonances, 
the spectra oscillate rapidly in this region and diminish gradually at higher photon energies.

\subsection{Band structures of MX$_2$ Trilayers}

\begin{figure}
\includegraphics[width=8cm]{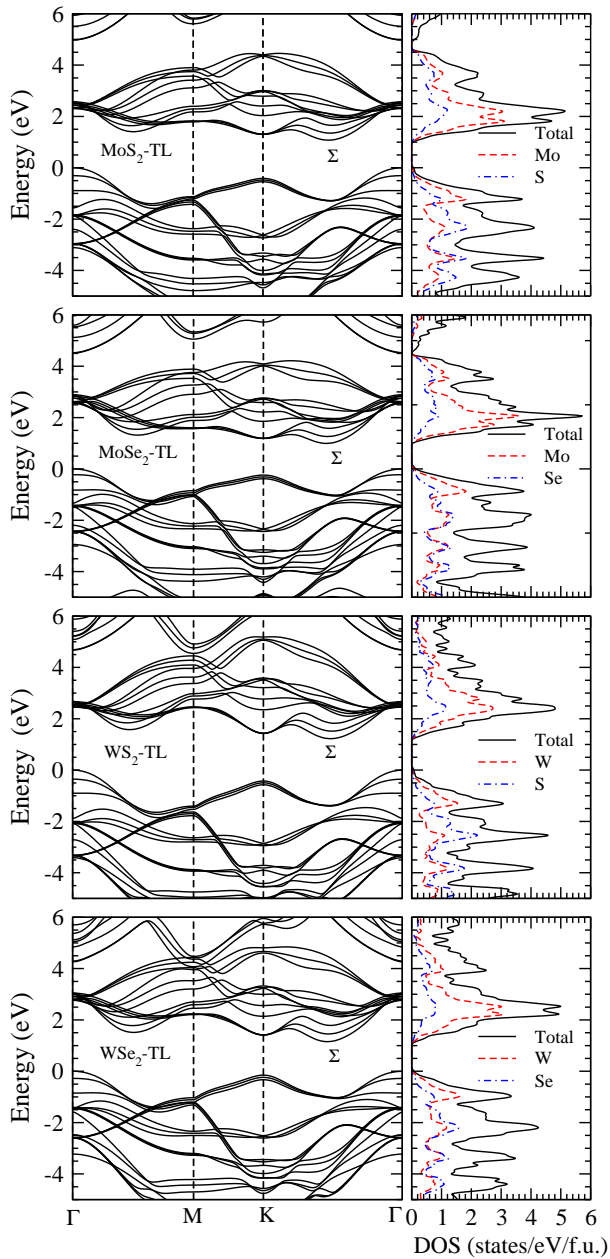}
\caption{\label{fig6}
Calculated band structures (left panels) and density of states (right panels) of the XM$_2$ TLs.
All the four materials exhibit an indirect band gap from the $\Gamma$ to $\Sigma$ point.
The top of the valence band is at 0 eV.}
\end{figure}

In order to investigate the effects of the interlayer interaction on the optical properties of the 
MX$_2$ multilayers, we also calculate the electronic structure as well as linear and nonlinear 
optical properties of the MX$_2$ TLs. The calculated band structures of the MX$_2$ TLs are 
shown in Fig. 6. If there were no interlayer interaction, the band structure of a MX$_2$ TL 
should be identical to that of the corresponding ML, except that the energy bands are now 
three fold degenerate. Nevertheless, the degenerated energy bands in the MX$_2$ TL are split 
due to the weak interlayer interaction. Overall, the band structure of the MX$_2$ TL is similar 
to that of the MX$_2$ ML except that the number of bands is tripled (see Figs. 3 and 6), 
because the band splitings due to the interlayer interaction are generally not large (Fig. 6). 
This similarity is especially clear in the calculated density of states for the MX$_2$ ML 
and TL (see Figs. 3 and 6). Significantly, however, these band splittings due to the
interlayer interaction lower the conduction bands near the $\Sigma$ point along the K-$\Gamma$ symmetry line 
to below the bottom of the conduction band at the K point, and also raize the top valence band at the $\Gamma$
point to above the top of the valence bands at the K point (see Fig. 6). Therefore, all the four
MX$_2$ TLs are semiconductors with an indirect band gap. And the band gaps in the MX$_2$ TLs are smaller 
than those of the MX$_2$ MLs by as much as 0.7 eV (see Table II). 
Interestingly, the direct energy gap 
at the K point is hardly affected by the interlayer interaction, and it decreases only slightly 
(within 0.1 eV) from the MLs to TLs (Table II). 
We note that the band structures of the MX$_2$ TLs shown in Fig. 6 are very similar to the 
band structures of the MX$_2$ multilayers reported before~\cite{aku12}.
The features that distinguish the MoX$_2$ MLs from the WX$_2$ MLs are still present
in the MX$_2$ TLs. For example, there is a small energy gap at about 5 eV in the MoX$_2$ 
TLs which is absent in the WX$_2$ TLs (Fig. 6).

\subsection{Second-order nonlinear optical susceptibility of MX$_2$ Trilayers}

As for the MX$_2$ MLs, we calculate the linear and nonlinear optical properties of the MX$_2$ TLs
by using both the GGA and scissors corrected band structures. Since there is no reported measurement on
the direct energy gap ($E_K$) at the K point of the MX$_2$ TLs and also the calculated
$E_K$ for bulk MX$_2$ and  the MX$_2$ TLs are close (Table II), we simply use the differences ($\Delta E_K^{bulk}$)
between the measured and calculated $E_K$ for bulk MX$_2$ for the scissors corrections. As mentioned before, 
the line shapes of the SHG susceptibility and dielectric function from the two types of calculations
are nearly identical, although the magnitude of these optical quantities from the scissors correction calculations
is reduced by about 25 \% (see Table III) and the peak positions are shifted upwards by about $\Delta E_K^{bulk}$.
Therefore, here we display only the spectra of the SH susceptibilities and dielectric functions 
obtained from the scissors corrected band structures in Fig. 7. 

\begin{figure*}
\includegraphics[width=16cm]{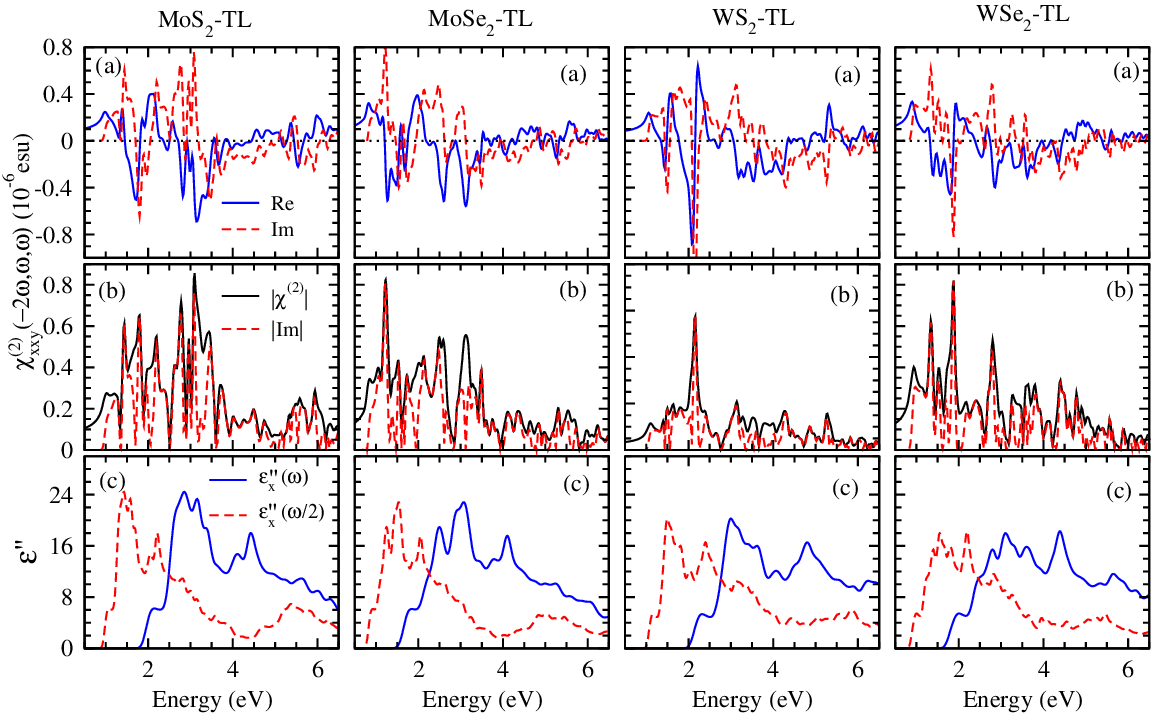}
\caption{\label{fig7}
(a) Real and imaginary parts as well as (b) the absolute value of the imaginary
part of the second-order susceptibility $\chi^{(2)}_{xxy}$ of the MX$_2$ trilayers calculated
from the scissors corrected band structures.
(c) Imaginary part $\varepsilon''$ of the corresponding dielectric function of the MX$_2$ trilayers.}
\end{figure*}

All the MX$_2$ multilayers with an odd number of MLs belong to the D$_{3h}$ symmetry class. 
Therefore, their nonzero elements of the SH susceptibility tensor are the same as that of the MX$_2$ MLs.
The results of our nonlinear optical calculations are consistent with this symmetry consideration. 
Figure 6 shows that the SH generation coefficients $\chi^{(2)}_{xxy}(-2\omega,\omega,\omega)$ are still significant
in the entire optical frequency range. Figure 6 also indicates that, as in the case of the MX$_2$ MLs,
the absorption edge of $\chi^{(2)}({\omega})$ and $\varepsilon''({\omega/2})$ is at half of the energy gap
at the K point for the MX$_2$ TLs, due to the two photon inter-band transitions at 
the K point. Therefore, the $\chi^{(2)}({\omega})$ of the MX$_2$ TLs are purely dispersive (i.e., lossless) 
for photon energy below half of the energy gap at the K point. We also find that the absorption 
edge of the imaginary part of the $\varepsilon''({\omega})$ is equal to the energy gap at the K 
point due to one photon inter-band transition.
Both single and double photon resonances occur above the energy gap at the K point 
in the MX$_2$ trilayers, resulting in rapid oscillations in the $\chi^{(2)}$ spectra which gradually
diminish in the high photon energy region.

Figures 5 and 7 clearly show that the SH susceptibilities of the MX$_2$ TLs are generally 
smaller than that of the MX$_2$ MLs, although their line shapes look rather similar.
In fact, if there were no interlayer interaction, the SH susceptibility of a MX$_2$ TL
would be 1/3 of that of the corresponding MX$_2$ ML. This is because the contributions from two MLs 
in the TL would cancel each other but the effective unit cell volume were tripled.
Table III shows that the ratio of the static values of the SH generation coefficients between the TLs and 
MLs varies in the range of 0.28$-$0.35, being indeed close to 1/3.
The slight deviations from 1/3 are due to the weak interlayer interaction.
For the incident laser beams with a wavelength of 810 nm wavelength (or 1.532 eV photon energy), 
the ratio can deviate more significantly from 1/3 and it is especially so for 
the WSe$_2$ ML and TL (Table III).
This may be expected because 1.532 eV falls within the regime of mixed single and double photon resonances
where not only the magnitude of the SH susceptibility gets reduced but also
the energy positions of the peaks shift as one moves from the ML to TL (see Figs. 5 and 7).


\subsection{Comparison with previous theoretical calculations and experiments}

Theoretical calculations of SH generation in the MoS$_2$ ML using an {\it ab initio} 
real-time approach\cite{Gru14} and also a semi-empirical tight-binding method\cite{Tro14} have
been reported recently. The $|\chi^{(2)}|$ spectra calculated previously within the IPA\cite{Gru14,Tro14}
are reproduced in Fig. 8(a) for comparison with the present calculation.
Figure 8(a) indicates that the $|\chi^{(2)}|$ spectra from the previous\cite{Gru14} and present
 {\it ab initio} IPA calculations agree quite well especially for the photon energy above 1.8 eV.
Below 1.8 eV, the $|\chi^{(2)}|$ spectrum from the present calculation has a much larger magnitude and
also have an additional peak located at $\sim$ 1.7 eV. The much broad features in the $|\chi^{(2)}|$ spectrum
from the previous {\it ab initio} calculations could be caused by much fewer $k$-points 
(a $k$-point mesh of $21\times21\times1$) and a larger broadening of 0.2 eV used in Ref. \onlinecite{Gru14}.
The $|\chi^{(2)}|$ spectrum from the previous tight-binding calculation\cite{Tro14} also
agrees rather with the present calculation except that below 1.3 eV. The much sharper features
in this previous calculation\cite{Tro14} could be due to a much smaller broadening of 0.03 eV used there.

The $|\chi^{(2)}|$ spectra for the MoS$_2$ ML calculated previously with electron-hole interaction 
taken into account\cite{Gru14,Tro14} are reproduced in Fig. 8(b). Figure 8(b) shows that the $|\chi^{(2)}|$ 
spectra from the previous real-time approach\cite{Gru14} and present IPA+SCI calculations agree rather well 
in both shape and magnitude except that below 1.4 eV. Note that the $|\chi^{(2)}|$ spectra from
the previous {\it ab initio} calculations\cite{Gru14} without and with inclusion of electron-hole interaction
look quite similar [see Figs. 8(a) and 8(b)]. This suggests that the pronounced excitonic peaks in the
linear optical spectra such as optical absorption\cite{Qiu13} might have largely been washed out in
the second-order nonlinear optical susceptibility, being consistent with the result in Ref. \onlinecite{Tro14}. 
The $|\chi^{(2)}|$ spectrum from the previous
tight-binding calculation with the excitonic effect included would also look quite similar to the
present IPA+SCI calculation if the present $|\chi^{(2)}|$ spectrum is blue-shifted by about 0.2 eV.
However, the magnitudes of the two spectra differ by about five times. 
Nevertheless, this five-fold increase in the magnitude of $|\chi^{(2)}|$ due to the inclusion of 
electron-hole interaction\cite{Tro14} 
is not seen in the previous {\it ab initio} real-time approach\cite{Gru14}.

Three groups recently reported observation of SH generation in the MoS$_2$ multilayers.\cite{nku13,mal13,yil13}
SH generation in WS$_2$ and WSe$_2$  multilayers were also reported\cite{zen13,Wan14b}.
One of the experiments~\cite{nku13} reported that, at 810 nm wavelength of Ti:sapphire laser, 
the $|\chi^{(2)}|$ of mechanically exfoliated MoS$_2$ ML is as large as $\sim$10$^{5}$ pm/V and for triangular 
flakes of the MoS$_2$ ML fabricated by chemical vapor deposition is $5\times10^{3}$ pm/V.~\cite{nku13}
The former value is about 170 times larger than our SCI theoretical $|\chi^{(2)}|$ value 
of $\sim$573 pm/V and the latter value is also about 9 times larger than our SCI value (Table III). 
In contrast, in another recent experiment\cite{yil13}, the $|\chi^{(2)}|$ value of the MoS$_2$ ML 
measured at 810 nm wavelength is about 320 pm/V, being about half of our SCI value (Table III).
In a more recent experiment~\cite{mal13}, the SH generations of the MoS$_2$ ML and TL were 
measured for a range of photon energy, and these 
experimental spectra are plotted in Fig. 8, together with the theoretical results.
Figure 8(a) indicates that the experimental $|\chi^{(2)}|$ spectrum for the MoS$_2$ ML has a line shape 
that agrees rather well with all three IPA theoretical $|\chi^{(2)}|$ spectra, {\it albeit} with a much smaller
magnitude. The experimental $|\chi^{(2)}|$ spectrum also agrees well in shape with the result of the tight-binding
calculation that included the electron-hole interaction\cite{Tro14}, although the peak in the experimental
spectrum appears to be red-shifted by $\sim$0.1 eV relative to the present SCI calculation and
also previous {\it ab initio} real-time approach to the excitonic effect\cite{Gru14}.
However, again, the experimental $|\chi^{(2)}|$ spectrum has a much smaller magnitude.
For example, the experimental $|\chi^{(2)}|$ value at 1.532 eV is $\sim$82 pm/V, 
being about seven (ten) times smaller than our theoretical SCI (IPA) value (Table III).
Finally, Fig. 8(c) indicates that the experimental $|\chi^{(2)}|$ spectrum for the MoS$_2$ TL roughly agrees in shape
the present SCI spectrum.

The fact that the measured $|\chi^{(2)}|$ values at 1.532 eV photon energy vary as much as three orders of magnitude, 
indicates the difficulties in accurate experimental deductions of SH generation coefficients of the MX$_2$ MLs
which depend on a number of experimental parameters\cite{nku13}.
On the other hand, we note that 1.532 eV photon energy falls in the energy range of mixed single and double photon resonances,
as mentioned before, and the magnitude of $|\chi^{(2)}|$ can change as much as two orders of magnitude in this region (Fig. 8).
Worse still, 1.532 eV photon energy is close to the energy position of a sharp peak in the $|\chi^{(2)}|$ spectra (Fig. 8).
Consequently, the peak position of different samples prepared by different methods could vary, and this variation
of the peak position could give rise to very different measured $|\chi^{(2)}|$ values at 810 nm wavelength.
Obviously, it would be helpful if the $|\chi^{(2)}|$ spectra are measured over a range of photon energy
and then are compared with each other and also with the theoretical results.
These large discrepancies between the experiments and also between the experiments and theoretical results
suggest that further experiments on these interesting ML materials would be desirable.

\begin{figure}
\includegraphics[width=7cm]{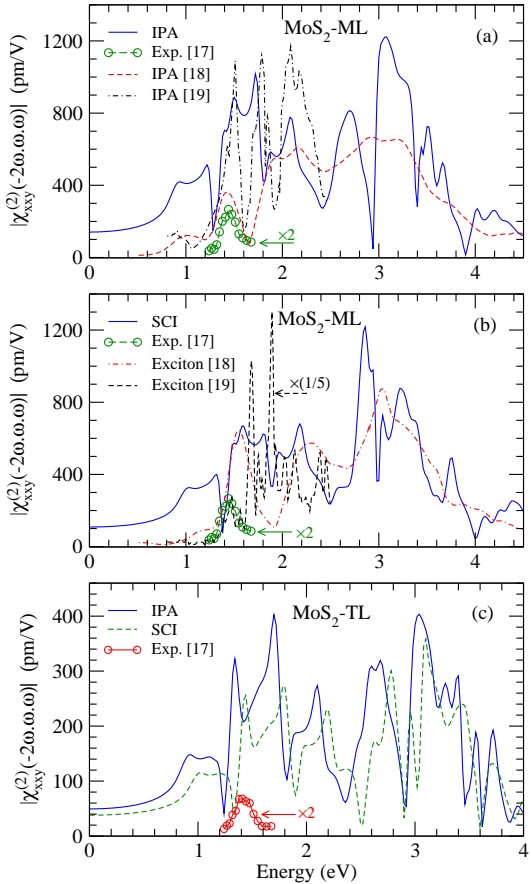}
\caption{\label{fig8}
Theoretical SH susceptibility $|\chi^{(2)}_{xxy}|$ of the MoS$_2$ ML from the present
and previous IPA calculations\cite{Gru14,Tro14} (a) and also from the present SCI calculation and previous 
{\it ab initio}\cite{Gru14} and tight-binding\cite{Tro14} calculations that included 
the excitonic effect (b). In (c), the $|\chi^{(2)}_{xxy}|$ of the MoS$_2$ TL from the present IPA and SCI
calculations are displayed. The experimental $|\chi^{(2)}_{xxy}|$ of the MoS$_2$ ML and TL
from Ref. \onlinecite{mal13} are also reproduced in (a-b) and (c), respectively. 
}
\end{figure}

As mentioned before, the SH susceptibility of the MX$_2$ TLs is generally reduced with respect to that of the MX$_2$ MLs.
This reduction factor should be 1/3 if there were no interlayer interaction. At 810 nm wavelength, the calculated ratio
of the SH susceptibility of the MoS$_2$ TL to the ML (0.28) deviates slightly from 1/3 but agrees rather well with
the experimental values of 0.25 and 0.21 (Table III). The pronounced deviation of the ratio from 1/3 is predicted to
occur in the WSe$_2$ ML and TL and is consistent with the experimental results\cite{zen13} (see Table III).
However, the measured ratio (0.60) between the WS$_2$ TL and the ML is significantly larger than 
the theoretical result (0.39) (Table III).
Nonetheless, given that the experimental $\chi^{(2)}$ values at this wavelength could vary a couple 
of orders of magnitude (Table III),
we believe that this level of the agreement in the $\chi^{(2)}$ ratio between the experiments 
and our theoretical results is rather satisfactory.
 
\section{Summary}
We have carried a systematic {\it ab initio} investigation of the second-order nonlinear optical properties of
the MX$_2$ (M$=$Mo,W and X$=$S, Se) MLs and TLs within the GGA plus scissors correction. We have used 
the accurate full-potential PAW method.
We find that the second-order nonlinear optical susceptibility [$\chi^{(2)}$]
of the MX$_2$ MLs in the entire optical photon energy range are large, being comparable to that of GaAs.
The calculated linear electro-optical coefficients in the low photon energy limit are also significant.
This shows that the four two-dimensional MX$_2$ semiconductors have promising potentials in, e.g., 
ultrathin second-harmonic and sum frequency generation devices,
electro-optical switches and light signal modulators.
The $\chi^{(2)}$ spectra of the MX$_2$ TLs are similar to the corresponding MX$_2$ MLs,
{\it albeit} with the magnitude reduced roughly by a factor of 3. The minor deviations from the 1/3 ratio  
are caused by the weak interlayer coupling of the electronic states. 
The prominent features in the calculated $\chi^{(2)}$ spectra of the MX$_2$ multilayers
have been successfully correlated with the peaks in the imaginary part of the corresponding optical dielectric function
$\varepsilon(\omega)$ in terms of single and double photon resonances.
The theoretical $\chi^{(2)}$ spectra of the MX$_2$ multilayers
are compared with the available experimental data, and the large discrepancies of as much as three orders of magnitude
among the measured $\chi^{(2)}$ data are analysized in terms of the present theoretical results. 
We hope that this work will stimulate further experimental investigations into the
second-order nonlinear optical responses and related properties of these fascinating few-layer MX$_2$ ultrathin films.

\section{Acknowledgments}

The authors thank Ana Maria de Paula for sending us her experimental second-harmonic generation data\cite{mal13} 
which are plotted in Fig. 8 and Thomas Pedersen for bringing his recent paper (Ref. \onlinecite{Tro14}) to our attention
as well as Claudio Attaccalite and Myrta Gr\"{u}ning for their communications on their recent theoretical 
results (Ref. \onlinecite{Gru14}). The authors gratefully acknowledge financial supports from the Ministry of Science and Technology, 
the Academia Sinica Thematic Research Program and the National Center for Theoretical Sciences of Taiwan.

\end{document}